\documentclass[12pt,preprint]{aastex}

\usepackage{amsmath}
\usepackage{amssymb}
\usepackage{bm}

\shorttitle{Heliospheric Jets}
\shortauthors{Drake et al}

\begin{document}


\title{A Model of the Heliosphere with Jets}


\author{
J.~F.~Drake\altaffilmark{1}, 
M.~Swisdak\altaffilmark{1}, 
M.~Opher\altaffilmark{2}
}


\altaffiltext{1}{University of Maryland, College Park, MD, USA;
 drake@umd.edu, swisdak@umd.edu}
\altaffiltext{2}{Astronomy Department, Boston
University, Boston, MA, USA ; mopher@bu.edu}


\begin{abstract}
An analytic model of the heliosheath (HS) between the termination
shock (TS) and the heliopause (HP) is developed in the limit in which
the interstellar flow and magnetic field are neglected. The
heliosphere in this limit is axisymmetric and the overall structure of
the HS and HP are controlled by the solar magnetic field even in the
limit in which the ratio of the plasma to magnetic field pressure,
$\beta=8\pi P/B^2$, in the HS is large. The tension of the solar
magnetic field produces a drop in the total pressure between the TS
and the HP. This same pressure drop accelerates the plasma flow
downstream of the TS into the North and South directions to form two
collimated jets. The radii of these jets are controlled by the flow
through the TS and the acceleration of this flow by the magnetic
field -- a stronger solar magnetic field boosts the velocity of the
jets and reduces the radii of the jets and the HP. Magnetohydrodynamic
(MHD) simulations of the global helioshere embedded in a stationary
interstellar medium match well with the analytic model. The results
suggest that mechanisms that reduce the HS plasma pressure downstream
of the TS can enhance the jet outflow velocity and reduce the HP
radius to values more consistent with the Voyager 1 observations than
in current global models.

\end{abstract}




\section{INTRODUCTION}\label{intro}
The historically accepted shape of the heliosphere is that of a
comet-like object with a long tail that is dragged downstream by the
flow of the local interstellar medium (LISM) past the sun
\citep{Parker61,Baranov93}. These early pictures, however, were based
on a hydrodynamic description of the solar outflow -- the solar
magnetic field was assumed to play a negligible role in the overall
structure of the heliosphere and its interaction with the solar
wind. Computational models based on the magnetohydrodynamic (MHD)
equations included the solar magnetic field as well as that of the
interstellar medium and also produced a heliosphere with a comet-like shape
\citep{Opher06,Pogorelov07,Opher09,Washimi11,Pogorelov13,Opher13}.

On the other hand, the measurements of energetic neutral atoms (ENAs)
by IBEX and CASSINI produced some surprises. These ENAs travel long
distances through the heliosphere without being influenced by the
ambient magnetic field and therefore yield information about the
large-scale structure of the heliosphere. The CASSINI ENA fluxes from
the direction of the nose and the tail were comparable, leading the
CASSINI observers to conclude that the heliosphere was ``tailless''
\citep{Krimigis09,Dialynas13}. The IBEX observations from the tail
revealed that the hardest spectrum of ENAs were localized in two lobes
at high latitude while the softest spectra were at low latitudes
\citep{McComas13}.

Recent MHD simulations using a monopole model for the solar magnetic
field, designed to reduce the numerical dissipation of magnetic energy
that arises from a conventional dipole model, revealed that the solar
magnetic field was strong enough to collimate the solar wind into a
pair of jets that flow to the North and South \citep{Opher15}. These
jets bend in the direction of the tail, pushed by the flow of the
LISM. The interstellar rather than the solar wind plasma flows between
these jets in the equatorial region downstream. Such
bent jets have been seen in protostellar systems
\citep{Fendt98,Gueth99} and clusters of galaxies
\citep{Owen76}. Astrophysical jets around massive black holes are
thought to be driven by centrifugal forces that sling the plasma along
a rotating helical magnetic field \citep{Blandford82}. However, the
jets in the case of the heliosphere are driven in the region
downstream of the TS as was proposed for
the Crab Nebula \citep{Begelman92,Chevalier94,Lyubarsky02}. In this
region of subsonic flow, the magnetic tension (hoop) force is strong
enough to collimate and drive the wind.

MHD models of the global heliosphere are complex and the mechanisms
that control the shape of the HP, the thickness of the HS, the
structure of the heliospheric jets, including the driver for the
outflow, remain uncertain. The Voyager 1 observations
have revealed that the thickness of the HS is around $30AU$,
which is substantially thinner than expected from the global
simulations. We present an analytic
model of the heliosphere outside of a spherically symmetric
TS where we neglect the ambient flow and magnetic field of the
LISM. Taking the resulting heliosphere as axisymmetric and
the flows within the HS as subsonic, we obtain the pressure
and magnetic field structure of the HS along with the radius $r_{hp}$
of the HP. The overall shape of the HS takes the classic form of an
astrophysical jet: the flows through the TS are accelerated to the
North and South by the solar magnetic field. The heliopause radius is
determined by continuity: the plasma flow through the TS must balance
the outflow through the jets. We present parallel global MHD
simulations in the limit of zero magnetic field and flow in the LISM
which support the analytic model.

One reason the influence of the solar magnetic field on the structure
of the heliosphere is often neglected in the literature is because the
pressure of the ambient plasma is large compared with that of the
magnetic field -- $\beta=8\pi nT/B^2\sim 10$ just downstream of the
TS. We show, however, that the total plasma pressure does not control
either the flows in or the thickness of the HS. The overall pressure
in the HS is balanced by the pressure in the LISM. It is the tension
force of the HS magnetic field that controls the pressure difference
between the TS and HP \citep{Axford72}. To the North and South there
is no tension force and this same pressure difference drives the axial
flow of the heliospheric jets. Thus, it is ultimately the solar
magnetic field that controls the large-scale structure of the HS.
\section{Analytic Model of the Heliosheath and Heliopause}
We consider a simple axisymmetric system in which there is no
LISM flow or magnetic field and the LISM is specified by its ambient
pressure $P_{LISM}$. We write down the steady-state MHD equations, including continuity, pressure, momentum and magnetic field,
\begin{equation}
{\bf \nabla}\cdot n{\bf V}=0,
\label{continuity}
\end{equation}
\begin{equation}
{\bf \nabla}\cdot P^{1/\Gamma}{\bf V}=0,
\label{pressure}
\end{equation}
\begin{equation}
M{\bf \nabla}\cdot n{\bf VV}=-{\bf\nabla}\left( P+\frac{B^2}{8\pi}\right)-\frac{B^2}{4\pi r}{\bf\nabla}r,
\label{momentum}
\end{equation}
\begin{equation}
{\bf \nabla}\times ({\bf V}\times{\bf B})=0, 
\label{faraday}
\end{equation}
where $r$ is the radius in cylindrical coordinates, $\Gamma$ is the
ratio of specific heats and ${\bf B}$ is in the azimuthal
direction. These equations are solved in the HS with boundary
conditions on the density, pressure, magnetic field and flow given
just downstream of the spherical TS which has a spherical coordinate
radius $R_s$. At the TS we assume that the azimuth flow $V_\phi$ is
zero and since there are no forces in the $\phi$ direction
(Eq.~(\ref{momentum})), we can take $V_\phi=0$ everywhere. Thus, from
Eq.~(\ref{continuity}) we can write
\begin{equation}
n{\bf V}={\bf\nabla}\phi\times{\bf\nabla}\psi ,
\label{nv}
\end{equation}
where $\psi$ is the stream function for the particle flux. Taking ${\bf B}=rB{\bf\nabla}\phi$ with ${\bf V}$ from Eq.~(\ref{nv}), we can reform Faraday's law in Eq.~(\ref{faraday}) as
\begin{equation}
{\bf\nabla}\psi\times{\bf\nabla}\left( \frac{B}{nr}\right)=0.
\end{equation}
This equation has components only in the $\phi$ direction so taking the dot product of this equation with ${\bf\nabla}\phi$ yields the constraint that
 $B/nr$ is constant along streamlines or
\begin{equation}
B/nr=f(\psi),
\end{equation}
where the function $f$ is only a function of $\psi$. The form of
$f$ can be determined by the boundary conditions along the
TS. Downstream of the TS we take the flow to be normal to the shock,
in the radial (in spherical coordinates) direction with a constant
value $V_s$ with the density $n_s$ also constant. So, from
Eq.~(\ref{nv}) we find $\partial\psi /\partial\theta = -R_s^2n_sV_s\sin\theta$ or
\begin{equation}
\psi=n_sV_sR_s^2\cos\theta_s,
\label{psi_s}
\end{equation}
where $\theta_s$ is the polar angle in spherical
coordinates at the shock. Thus, the variation of $\psi$ along the shock is
known. Similarly, we know from the solutions of the Parker spiral
magnetic field that along the shock $B=B_s\sin\theta_s$ so
$B/nr=B_s/n_sR_s$ is a constant and so is $f=B_s/n_sR_s$. Throughout the HS we have 
\begin{equation}
\frac{B}{nr}=\frac{B_s}{n_sR_s}.
\end{equation}
Turning to the pressure equation and using Eq.~(\ref{nv}), we find 
${\bf\nabla}\phi\times{\bf\nabla}\psi\cdot{\bf\nabla}(P/(n^\Gamma)=0$
so $P/n^\Gamma$ is also constant along a streamline and therefore a function only of $\psi$. As before, we can evaluate it along the TS where it is given by $P_0/n_s^\Gamma$. Thus, 
\begin{equation}
P=P_s\left(\frac{n}{n_s}\right)^\Gamma =P_s\left(\frac{BR_s}{B_sr}\right)^\Gamma .
\label{P}
\end{equation}
Thus, both $P$ and $n$ in the HS are linked to $B$ and $r$. 

We now focus on the high $\beta$ limit, which is most relevant to the
HS, where the pressure associated with interstellar pickup ions
dominates the magnetic pressure. Specifically, we
take $\beta$ to be a large parameter. In addition, since the flows are
subsonic downstream of the TS, inertial forces are also small. We can
therefore write the plasma pressure in a series $P_0+P_1+...$ with
$P_1\sim B^2/8\pi\sim MnV^2/2\ll P_0$. Thus, to lowest order
Eq.~(\ref{momentum}) becomes
\begin{equation}
0=-{\bf\nabla}P_0(B,r),
\label{momentum0}
\end{equation}
where $P_0$ is an explicit function of $B$ and $r$ through
Eq.~(\ref{P}). Equation (\ref{P}) requires that to lowest order the
pressure in the HS is constant everywhere and is given the value
$P_s$ at the TS. The density is also a constant,
$n_s$. Since $P$ is linked to $B$ and $r$ through
Eq.~(\ref{P}) the constancy of $P_0$ requires that
\begin{equation}
B=B_s\frac{r}{R_s}
\label{B}
\end{equation}
so that $B$ increases with radius outside of the TS
\citep{Axford72,Chevalier94}. At first order we include the inertial
terms and magnetic field in the momentum equation, which becomes
\begin{equation}
Mn_s{\bf V}_0\cdot{\bf \nabla}{\bf V}_0=-{\bf\nabla}\left(
P_1+\frac{B_s^2r^2}{8\pi R_s^2}\right)-\frac{B_s^2r}{4\pi
  R_s^2}{\bf\nabla}r,
\label{momentum1}
\end{equation}
where $n$ has been replaced by $n_s$ and ${\bf V}_0={\bf\nabla}\phi\times{\bf\nabla}\psi/n_s$. 

\begin{figure}
\epsscale{.70}
\includegraphics[keepaspectratio,width=3.0in]{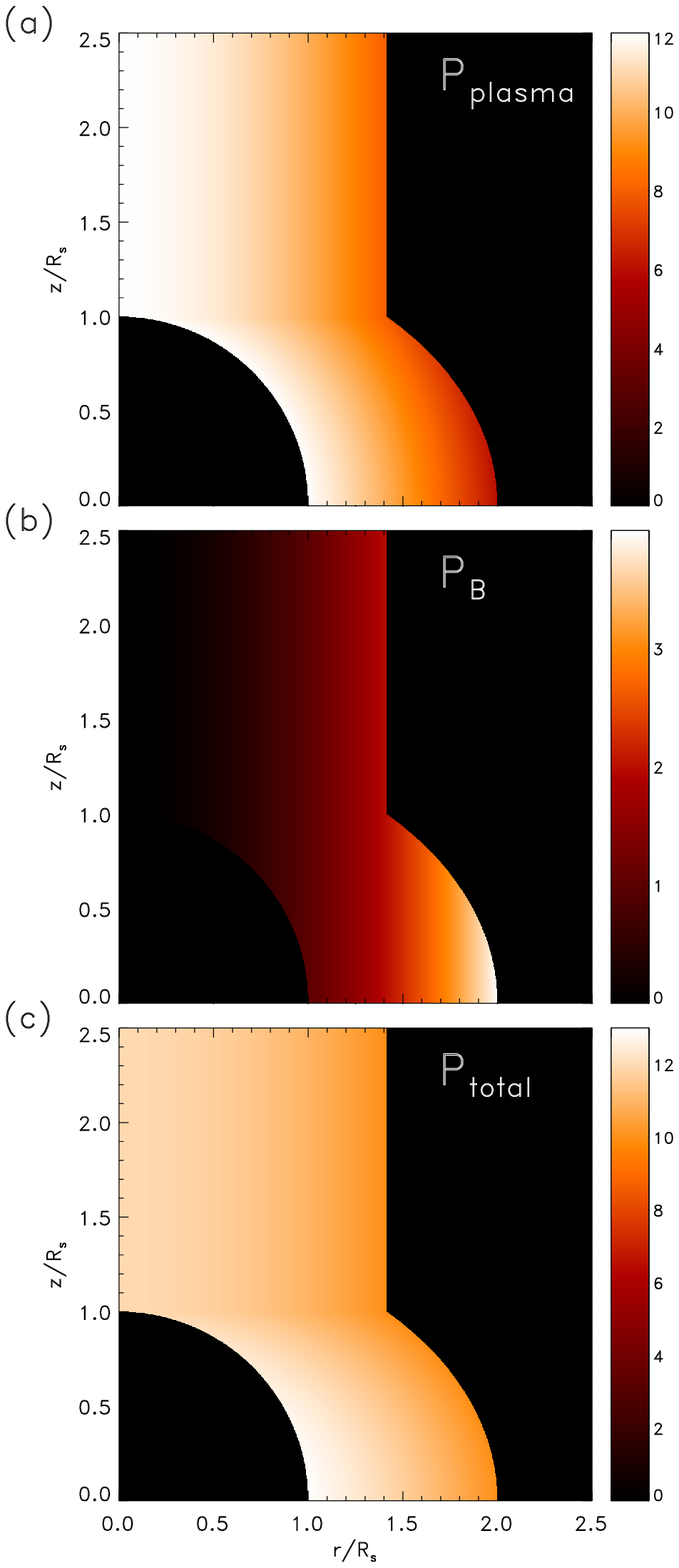}
\caption{\label{helio_model} 2-D images in the HS of the plasma pressure
  in (a), the magnetic pressure in (b) and the total pressure in (c),
  all normalized to $P_{Bs}=B_s^2/8\pi$, where $B_s$ is
  the magnetic field just downstream of the TS at the equator. The
  images are for $\Delta P=P_s-P_{LISM}=2B_s^2/8\pi$ and
  $\beta_s=P_s/P_{Bs}=12$, where $P_s$ is the plasma pressure
  downstream of the TS and $P_{LISM}$ is the pressure of the LISM.}
\end{figure}

Before discussing the flows in the HS, we consider the weak flow limit
of Eq.~(\ref{momentum1}) so that the inertial forces in the radial
direction can be discarded. In this limit the magnetic tension force
in Eq.~(\ref{momentum1}) causes the plasma pressure and total pressure
to decrease with radius. This limit is artificial for the HS since the
flow $V_s$ downstream of the TS is comparable to the Alfv\'en speed
and the associated radial inertial forces are comparable to the
magnetic forces. Nevertheless, this limit illustrates how the pressure
in the HS varies. We do not require zero pressure gradient in $z$. The
pressure drop from the TS to the LISM is balanced by magnetic tension
in the radial direction but the same pressure drop also develops from
the equator to the outflow jets to the North and South. This pressure
gradient along $z$ drives the outward flows associated with the
jets. Thus, we integrate Eq.~(\ref{momentum1}) from the TS outwards to
obtain an explicit expression for $P_1$,
\begin{equation}
P_1(r)=-\frac{B_s^2}{4\pi R_s^2}(r^2-R_s^2\sin^2\theta_s),
\label{P1}
\end{equation}
where $\theta_s(z)$ is the value of $\theta$ at the TS and is
dependent on $z$. Pressure balance across the HP, which requires that
$P(r_{hp})+B^2(r_{hp})/8\pi =P_{LISM}$, then yields an explicit
expression for $r_{hp}$,
\begin{equation}
\frac{r_{hp}^2}{R_s^2}=\frac{8\pi\Delta P}{B_s^2}+2\sin^2\theta_s,
\label{rhp}
\end{equation}
where $\Delta P=P_s-P_{LISM}$. At this stage in the calculation the
pressure difference $\Delta P$ remains undetermined. We will show,
however, that the requirement that the mass flow into the HS across
the TS balance that out the two jets constrains the pressure
difference. In Fig.~\ref{helio_model} we show 2-D plots of the plasma
pressure, the magnetic pressure and the total pressure in the $r$, $z$
plane in the HS. The data is shown for $8\pi\Delta P/B_s^2=2$, which
as shown later is the upper limit on $\Delta P$. The inner boundary of
the data shown is the TS and the outer boundary is the HP. The radius
of the HP peaks at the midplane and falls off with $z$ until $z>R_s$,
where it remains constant, forming the Northward jet. The plasma and
total pressure decrease with radius $r$ while the magnetic pressure
increases with $r$. The total pressure falls off with distance from
the equator until it approaches a constant value in the jet. We will
show that this pressure difference, which is a consequence of magnetic
tension, drives the jet outflow. Along the axis $P$ remains constant
at $P_s$. This shape was obtained by neglecting the radial plasma
inertia, an assumption which breaks down where the straight portion of
the HP in Fig.~\ref{helio_model} intersects the curved portion at
$z/R_s=1$. The sharp kink in the HP is not real and
is not seen in the MHD simulations discussed later.  In
Fig.~\ref{helio_cuts} cuts along $r$ of the total, plasma and magnetic
pressures from the data of Fig.~\ref{helio_model} are shown at the
equator in (a) and across the jet in (b).

We now discuss the flows driven in the HS. We derive a Bernoulli-like
equation by taking the dot product of Eq.~(\ref{momentum1}) with ${\bf
  V}_0$ and integrating along the streamline,
\begin{equation}
\frac{1}{2}Mn_sV_0^2+P_1+\frac{B_s^2r^2}{4\pi R_s^2}=\frac{1}{2}Mn_sV_s^2+\frac{B_s^2\sin^2\theta_s}{4\pi}.
\label{V0}
\end{equation}
Unfortunately, this equation does not enable us to calculate $V_0$
throughout the HS because it requires that we know the
trajectory of a stream line within the HS to link the local radius $r$
with its position at the TS where $\theta_s$ is known. Along the axis
where $\theta_s=0$ Eq.~(\ref{V0}) gives $V_0=V_s$, since
the pressure is constant and $B$ is zero. Similarly, the velocity at
the jet radius $r_{jet}$, can be calculated since at that location
$\theta_s=\pi/2$. We find $V_0^2(r_{jet})=V_s^2+c_{As}^2$, where
$c_{As}^2=B_s^2/4\pi Mn_s$. Thus, the increase in the jet velocity
above $V_s$ is linked to the Alfv\'en velocity based on the magnetic
field strength $B_s$ at the TS. More generally, we can calculate
$V_0(r)$ across the jet radius by noting that within the jet
\begin{equation}
V_0=V_{0z}=-\frac{1}{n_sr}\frac{\partial\psi}{\partial r}.
\end{equation}
From Eq.~(\ref{psi_s}) $\sin\theta_s$ can be written in terms of $\psi$ so we are left with a single equation for $\psi$ across the jet,
\begin{equation}
\frac{1}{n_s^2r^2}\left(\frac{\partial\psi}{\partial r}\right)^2=V_s^2+2c_{As}^2\left( 1-\frac{\psi^2}{n_s^2R_s^4V_s^2}\right),
\label{psi_jet}
\end{equation}
 where $\psi$ varies from $n_sV_sR_s^2$ at the jet axis to $0$ at the HP. The equality of the particle fluxes through the TS and jets requires that Eq.~(\ref{psi_jet}) produce the requisite jump in $\psi$ across the jet. Equation (\ref{psi_jet}) can be simplified by defining an angle variable $\cos\theta = \psi/n_sV_sn_s^2$,
\begin{equation}
\frac{R_s^4\sin^2\theta}{r^2}\left(\frac{\partial\theta}{\partial r}\right)^2=1+2\frac{c_{As}^2}{V_s^2}\sin^2\theta,
\label{theta_jet}
\end{equation}
where $\theta$ varies from $0$ at the jet axis to $\pi /2$ at the HP. This equation can be integrated directly to obtain the jet radius $r_{jet}$,
\begin{equation}
r_{jet}^2=2R_s^2\frac{\tan^{-1}(\sqrt{2}c_{As}/V_s)}{\sqrt{2}c_{As}/V_s}
\label{rjet}
\end{equation}
The jet radius is a maximum for $c_{As}\ll V_s$ when the jet outflow
velocity is given by $V_s$. In this limit the conservation of particle
flux reduces to the jet cross-sectional area being equal to the TS
area or $r_{jet}=\sqrt{2}R_s$. With increasing $c_{As}$ the outflow
velocity of the jet increases and $r_{jet}$ decreases. For
$c_{As}\gg V_s$, $r_{jet}\propto R_s\sqrt{V_s/c_{As}}$. An expression for the pressure jump $\Delta P$ between the TS and the
LISM can be calculated from Eq.~(\ref{rhp}), which is exact in the
jet where $V_{0r}=0$,
\begin{equation}
\Delta P=2\frac{B_s^2}{8\pi}\frac{\tan^{-1}(\sqrt{2}c_{As}/V_s)}{\sqrt{2}c_{As}/V_s}.
\label{deltaP}
\end{equation}
The pressure jump is a maximum when $c_{As}$ is small and
decreases with increasing $c_{As}$. The dependence of $r_{jet}$ and
$\Delta P$ are shown as functions of $c_{As}/V_s$ in
Fig.~\ref{rjet_model}. From Eq.~(\ref{rhp}) $r_{hp}$ therefore also decreases with increasing $c_{As}/V_s$.

\section{Global MHD Simulations}
We have carried out MHD simulations of the global heliosphere without
an interstellar wind and magnetic field. Our model is based on the 3D
multi-fluid MHD code BATS-R-US. It envolves one ionized and four
neutral H species as well as the magnetic field of the sun. We used a
monopole configuration for the solar magnetic field to eliminate
artificial reconnection across the heliospheric current sheet. The
basic parameters of the simulation are the same as those described in
Opher {\it et al.} 2015. The computational grid was $\pm 3000~AU$ in
each direction. Parameters of the solar wind at the inner boundary at
$30~AU$ were: $v_{SW} = 417 km/s$, $n_{SW} = 8.74 \times 10^{-3}
cm^{-3}$, $T_{SW} = 1.087 \times 10^{5} K$ and the Parker spiral
magnetic field with a radial component $B_{SW} = 7.17 \times 10^{-3}
nT$ at the equator (with an azimuthal component $B_\phi=0.22nT$). The
solar wind flow at the inner boundary is assumed to be spherically
symmetric and the magnetic axis is aligned with the solar rotation
axis. For the LISM we assume $T_{ISM}= 6519~K$ while the plasma
density was raised to $0.483/cm^3$ to make up for the absence of
pressure associated with the interstellar magnetic field. The number
density of H atoms in the interstellar medium is $n_{H} =
0.18~cm^{-3}$ and the temperature is the same as for the interstellar
plasma. The z-axis is parallel to the solar rotation axis. The grid
has cells ranging from $2.93~AU$ at the inner boundary to
$187.5~AU$ at the outer boundary. The simulation had a resolution of
$3.0AU$ between $z=\pm 750 AU$ and $x=\pm 305AU$; $y=\pm 400AU$,
encompassing the entire HS. The run was stepped forward for $3061$ years.

\begin{figure}
\epsscale{.70}
\includegraphics[keepaspectratio,width=3.0in]{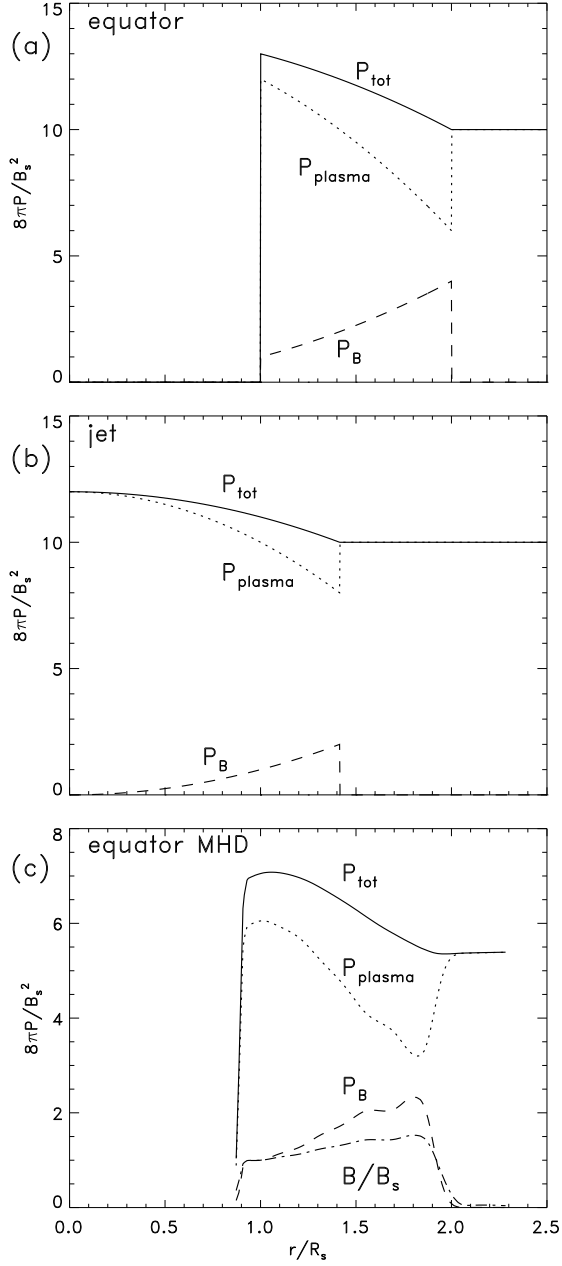}
\caption{\label{helio_cuts} Cuts through the data of
  Fig.~\ref{helio_model} of the total pressure (solid), plasma
  pressure (dotted) and magnetic pressure (dashed) versus $r$ at the
  equator in (a) and within the jet in (b). In (c) cuts from the MHD simulation of Fig.~\ref{mhd_hp} along the equator.}
\end{figure}

\begin{figure}
\epsscale{.70}
\includegraphics[keepaspectratio,width=6.0in]{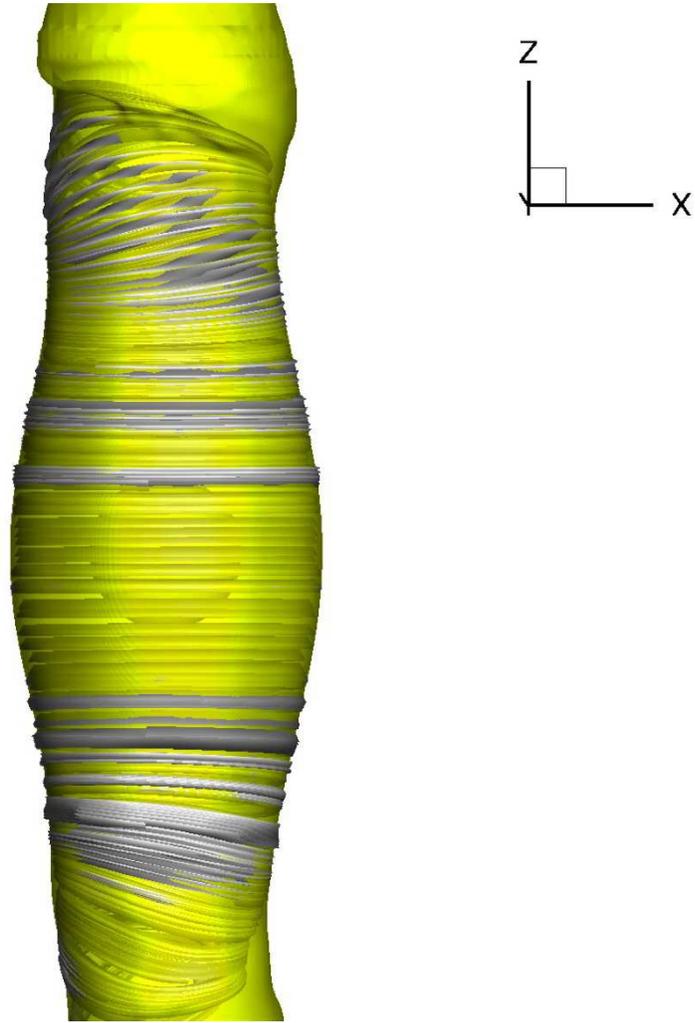}
\caption{\label{mhd_hp} The heliopause as defined by $\ln T = 13.9$ from an MHD simulation embedded in an ambient interstellar medium with no mean flow and zero magnetic field.  The gray lines are the solar magnetic field with the TS visible as a disc.}
\end{figure}

\begin{figure}
\epsscale{.20}
\includegraphics[keepaspectratio,width=3.0in]{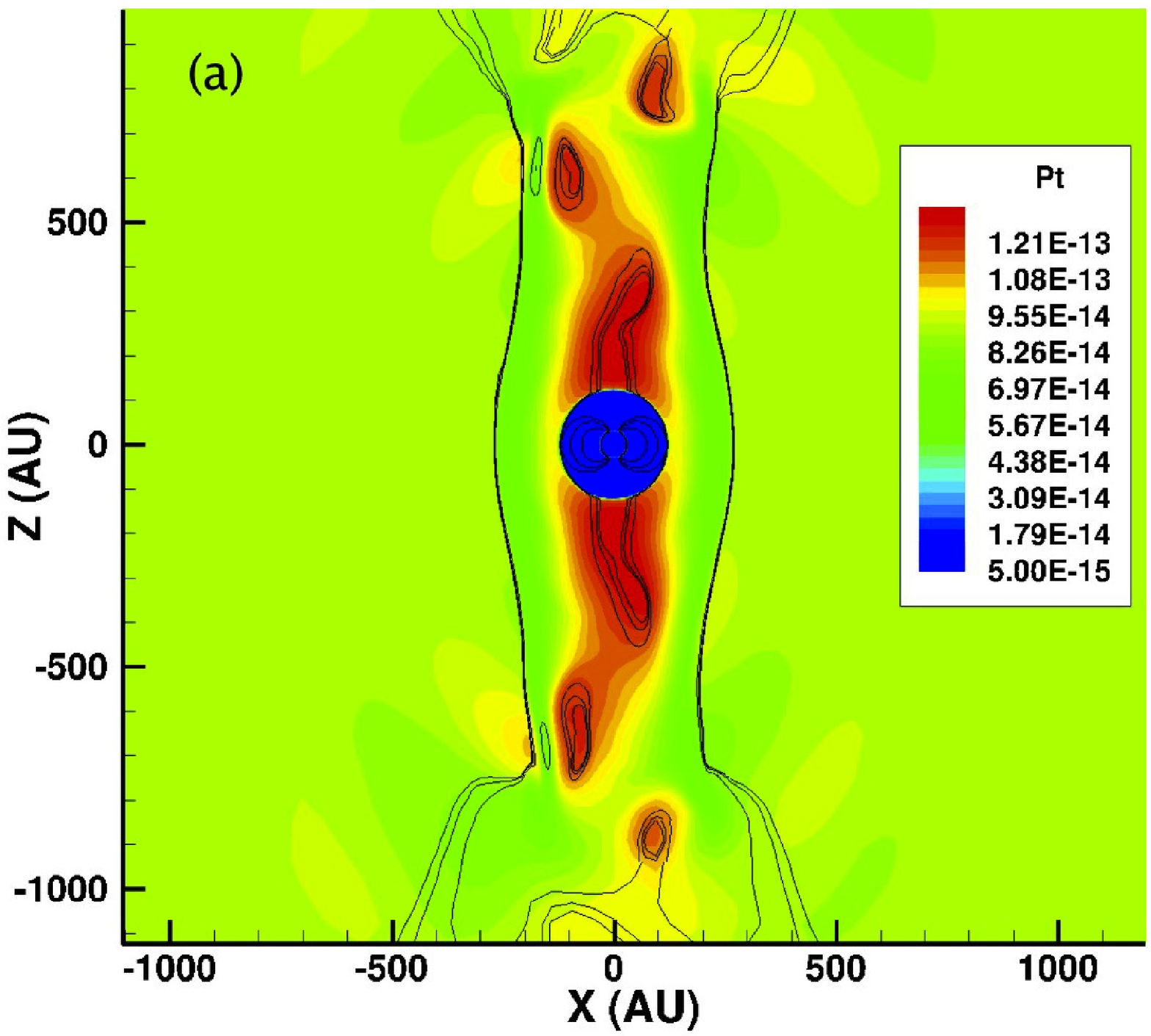}
\includegraphics[keepaspectratio,width=3.0in]{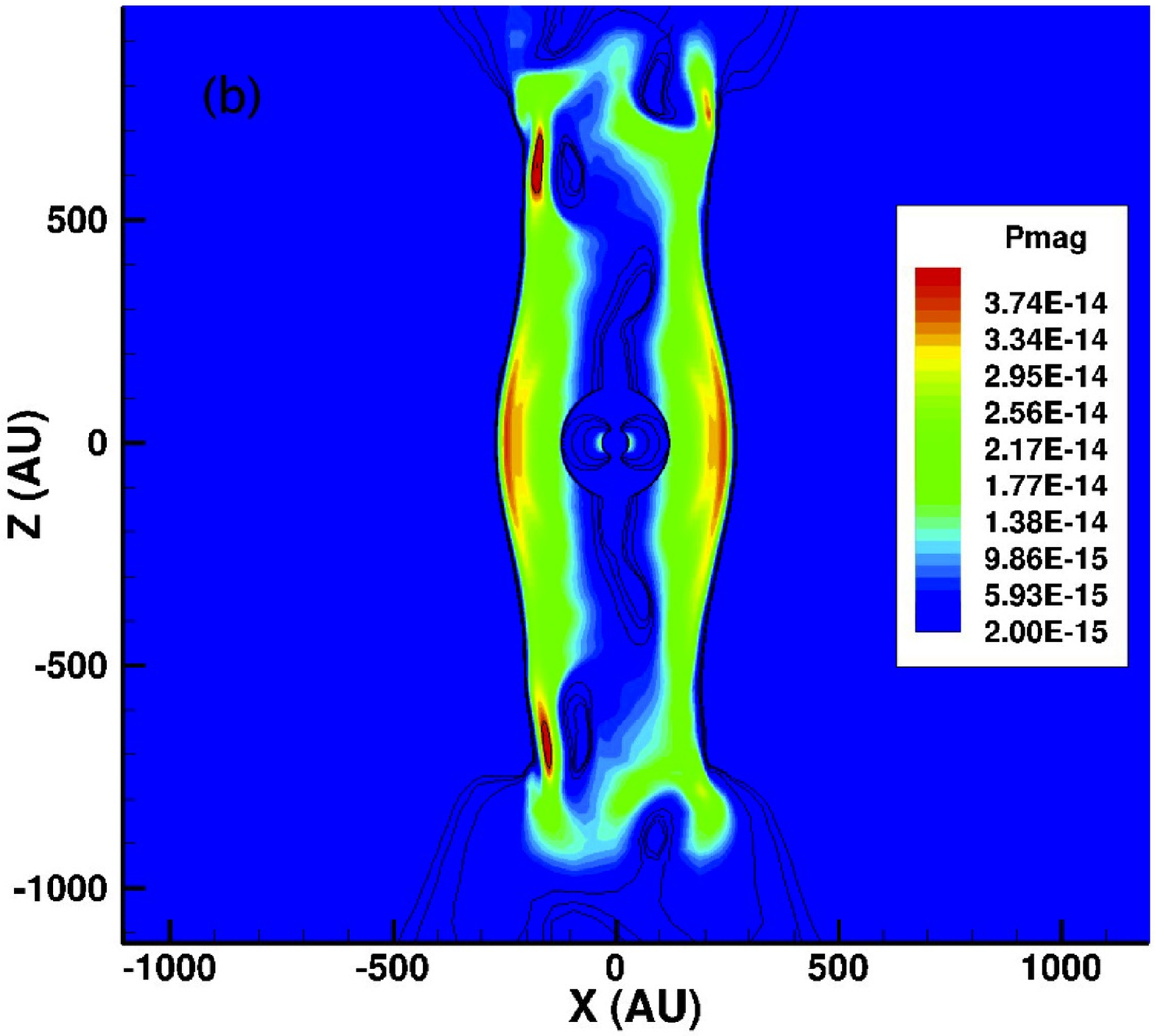}
\includegraphics[keepaspectratio,width=3.0in]{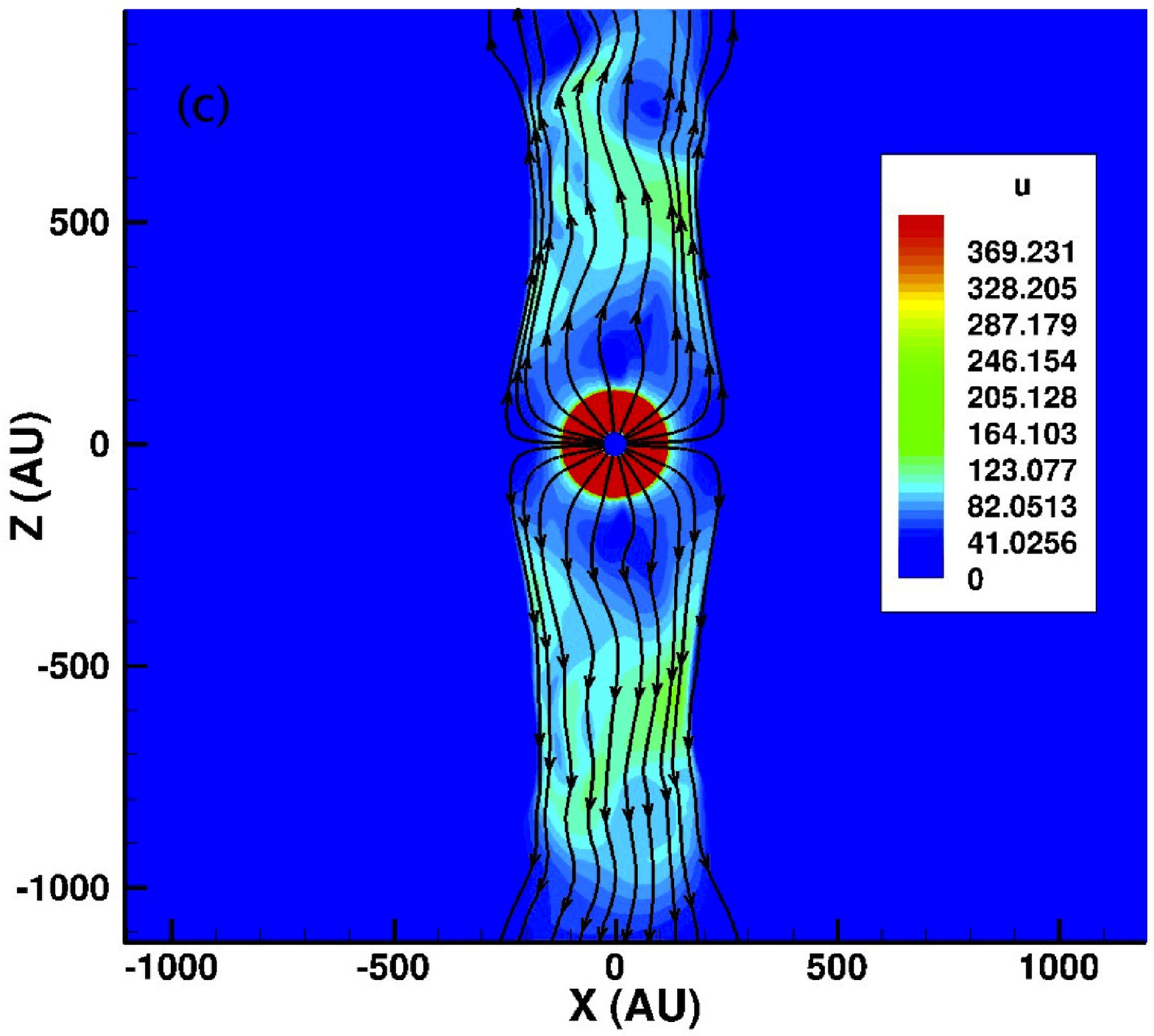}

\caption{\label{mhd_P} The plasma and magnetic pressures ($Pa$) in (a) and (b), and the plasma speed ($km/s$) and streamlines in (c). All in the $x-z$ plane through the center of the heliosphere from the simulation in Fig.~\ref{mhd_hp}. }
\end{figure}

In Fig.~\ref{mhd_hp} we show in yellow the surface of the HP as
defined by $ln T=13.9$. The simulation reveals jets to the North and
South as in the analytic model. The HP bulges at the equator as in the
model. The gray lines are the solar magnetic field. Shown in
Fig.~\ref{mhd_P} in the $x-z$ plane are the plasma pressure in (a),
the magnetic pressure in (b) and the speed and streamlines in (c). As
in the model, the plasma pressure decreases with cylindrical radius $r$ away
from the jet axis while the magnetic pressure increases with $r$ and
the strongest magnetic fields are in the equatorial region just
upstream of the HP. The streamlines reveal the North and South
directed outflows that make up the jets. The HP boundary
does not reveal the sharp indentation seen in the model. Finally, in
Fig.~\ref{helio_cuts}(c) we show cuts of the total pressure (solid),
the plasma pressure (dotted), the magnetic pressure (dashed) and the
magnetic field (dot-dashed) in cuts along $r$ at the equator from just
upstream of the TS to past the HP. The pressures have been normalized
to $B_s^2/8\pi$, $B$ to $B_s$ and $r$ to $R_s$ with $R_s=135AU$ taken to be
the location of the maximum of $P_{plasma}$. The cuts are in
remarkable agreement with the cuts from the model in (a). In the
simulation $c_{As}/V_s=1.1$, which from Fig.~\ref{rjet_model}, yields
$8\pi\Delta P/B_s^2=1.3$ compared with the measured value of $1.5$
from Fig.~\ref{helio_cuts}(c). For $8\pi\Delta P/B_s^2=1.5$
Eq.~(\ref{rhp}) yields $r_{hp}/R_s=1.9$ at the equator, essentially
identical to the HP radius in the cuts, and 
the ratio of the HP radius at the equator to that in the jet is $0.64$
(Eq.~(\ref{rhp})) compared with the reasured value of $0.66$. Finally
the measured particle flux through the TS is the same as that out the
jets.
\section{Discussion and Conclusions}
\begin{figure}
\epsscale{.70}
\includegraphics[keepaspectratio,width=4.0in]{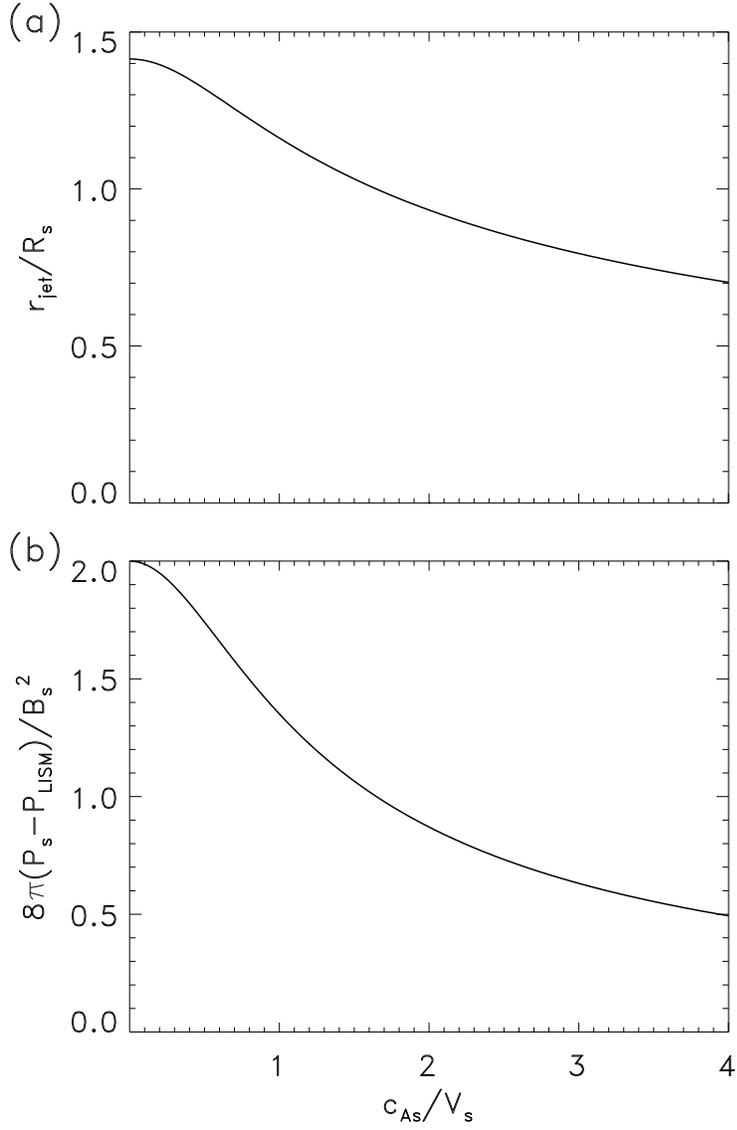}
\caption{\label{rjet_model} The jet radius $r_{jet}$ and the pressure jump between the TS and interstellar medium, $P_s-P_{LISM}$, versus the Alfv\'en speed downstream of the TS at the equator, $c_{As}$. }
\end{figure}

We have explored the structure of the HS and HP when the interstellar
flow and magnetic field are neglected and the system can be treated as
axisymmetric. We show that even in the limit in which $P\gg B^2/8\pi$
in the HS the magnetic field controls the large-scale
structure of the HS and drives Northward and Southward directed
jets. To lowest order the pressure in the HS is balanced by the
pressure in the interstellar medium. The magnetic field controls the
pressure variation within the HS and re-directs and boosts the flow
across the TS to the North and South to form heliospheric jets. The
radial distance from the TS to the HP and the jet radii are controlled
by the requirement that the plasma flowing into the HS across the TS
flows outwards in the jets (see also \cite{Yu74}). For very weak
magnetic fields the jet outflow velocity is the same as the velocity
$V_s$ downstream of the TS. In this limit the total cross-sectional
area of the jets is equal to the area of the TS and
$r_{jet}=\sqrt{2}R_s$. With increasing magnetic field strength the jet
outflow velocity increases and the radii of the HP and
the outflow jet decrease (Eq.~(\ref{rjet}) and
Fig.~\ref{rjet_model}(a).

The global MHD models of the heliosphere \citep{Malama06,Pogorelov07,Opher09}
produce HS thicknesses that are around $50-70AU$, substantially larger
than the value of $30AU$ determined from Voyager 1's
crossing of the HP in 2012 \citep{Stone13}. The results here suggest
that mechanisms that increase the jet outflows will reduce the HP
radius. Pressure reductions in the downstream region associated with
thermal conduction or other mechanisms might produce such enhanced flows.

There is evidence from both the present MHD simulations and those
carried out the earlier \citep{Opher15} that the jets are subject to
large-scale instabilities.  The resulting turbulence might be a driver
of anomalous cosmic rays \citep{Stone05,stone08}. High time-resolution
ENA measurements might be able to establish the existence of the
heliospheric jets and associated turbulence. For a jet radius of
around $140AU$ and an Alfv\'en velocity of around $100km/s$, the
Alfv\'en transit time is around $10$ years. The jet turbulence might,
of course, cascade to smaller scales so the relevant time scales could
be shorter.

\acknowledgments This work has been supported by NASA Grand Challenge
NNX14AIB0G and NASA awards NNX14AF42G, NNX13AE04G and NNX13AE04G. The
MHD simulations were carried out on Pleades at the NASA Ames Research
Center under the award SMD-14-4986. We acknowledge support
from the International Space Science Institute for the team ``Facing
the Most Pressing Challenges to Our Understanding of the Heliosheath
and its Outer Boundaries.''


\clearpage

\end{document}